      [1999/12/01 v1.4c Il Nuovo Cimento]
\documentclass{cimento}

\renewcommand{\d}{\mathrm{d}}

\renewcommand{\d}{\mathrm{d}}

\newcommand{\bea}{\begin{eqnarray}}
\newcommand{\eea}{\end{eqnarray}}

\newcommand{\nn}{\nonumber \\}
\newcommand{\nnn}{\nonumber }

\def\slash#1{\setbox0=\hbox{$#1$}  
   \dimen0=\wd0     
   \setbox1=\hbox{/} \dimen1=\wd1  
   \ifdim\dimen0>\dimen1   
      \rlap{\hbox to \dimen0{\hfil/\hfil}} 
      #1     
   \else     
      \rlap{\hbox to \dimen1{\hfil$#1$\hfil}} 
      /      
   \fi}      %

\usepackage{graphicx}  
\title{Model Prediction for the Transverse Single Target-Spin Asymmetry in inclusive DIS}
\author{M.~Schlegel\from{ins:x}
}
\instlist{\inst{ins:x} Institute for Theoretical Physics, University of T\"ubingen, Germany
}
\PACSes{\PACSit{}{13.60.Hb}
\PACSit{}{13.88.+e}}

\begin{document}

\maketitle

\begin{abstract}
The single-spin asymmetry of unpolarized leptons scattering deep-inelastically off transversely polarized nucleons is discussed. This observable is generated by a two-photon exchange between lepton and nucleon. In a partonic description of the asymmetry the non-perturbative part is given in terms of multiparton correlations: quark-gluon correlation functions and quark-photon correlation functions. Recently, a model for quark-gluon correlation functions was presented where these objects were expressed through non-valence light cone wave functions. Using this model, estimates for the single-spin asymmetries for a proton and a neutron are presented.
\end{abstract}

\section{Introduction}
One of the most fundamental and basic processes in hadronic physics is the deep-inelastic scattering (DIS) of leptons off nucleons, $\mathrm{l}(l)+\mathrm{N}(P)\to \mathrm{l}(l^\prime)+\mathrm{X}$. 
Single-spin observables in inclusive DIS with either the lepton or nucleon being transversely polarized strictly vanish due to time-reversal invariance for a single photon exchange~\cite{Christ:1966zz}. This argument fails if two (or more) photons are exchanged between lepton and nucleon. 

Experimentally, 
a recent measurement of the single-spin asymmetry (SSA) for a transversely polarized nucleon, denoted by $A_{UT}$, was performed by the HERMES collaboration~\cite{HERMES}, and again a result consistent with zero was found within an error of about $10^{-3}$. Interestingly, preliminary data taken from (ongoing) precision measurements of $A_{UT}$ at Jefferson Lab seem to indicate a non-zero effect~\cite{JLab}.

A theoretical description of the SSA $A_{UT}$ in a partonic picture needs to deal with two distinctive and complementary physical situations: The exchange of two photons between the lepton and either (i) one {\it single} quark or (ii) two {\it different} quarks.
The asymmetry has been studied in Refs.~\cite{OwnWork1, OwnWork2, SameQuark,DiffQuark} for massless quarks. It was found that this observable generically behaves like $M/Q$  where $M$ denotes the nucleon mass, $Q^2=-q^2$, and $q=l-l^\prime$ the 4-momentum transfer to the nucleon. Thus the asymmetry is a power suppressed ('twist-3') observable, and can be expressed in terms of multipartonic non-perturbative quark-gluon (scenario (i)) and quark-photon (scenario (ii) correlation functions. Effects of a finite quark mass proportional to the transversity distribution $h_1^q(x)$ are also relevant for scenario (i) and have been studied in Ref.~\cite{Afanasev}. 
%



\section{$A_{UT}$ in a partonic picture} The DIS differential cross section can be analyzed in terms of the commonly used DIS variables that are defined as $x_B=Q^2/(2P\cdot q)$ and $y=P\cdot q/P\cdot l$. 
For the description of $A_{UT}$ a transverse (to the lepton plane) spin vector $S_T$ of a polarized nucleon is needed. An azimuthal angle $\phi_s$ between $S_T$ and the lepton plane determines the spatial orientation of $S_T$. 
\\
An analysis of the SSA $A_{UT}$ in inclusive DIS in a partonic picture has to be performed at subleading twist accuracy \cite{OwnWork1,SameQuark,DiffQuark}. This requires the introduction of typical hadronic matrix elements of certain partonic operators that encode non-trivial correlations of the transverse nucleon spin and the transverse partonic motion~\cite{Baccetal}, as well as multipartonic correlations~\cite{ABMS,DiffQuark}. However, the effects of transverse partonic motion and multipartonic correlations are not independent. In fact, they can related to each other by means of the QCD-equation of motion (EOM)~\cite{Baccetal}. An additional dependence originates from the relation between the Sivers function and the so-called Qiu-Sterman matrix element~\cite{BMPQS}. If one applies the twist-3 factorization formalism of Ref.~\cite{DiagFact} to the SSA $A_{UT}$ all of these hadronic matrix elements are to be convoluted with corresponding partonic hard cross sections, and eventually summed up (cf.~\cite{SameQuark}). \\
The hard cross sections relevant for scenario (i) are calculated in perturbation theory to $\mathcal{O}(\alpha^3)$ to obtain a non-zero result. This includes interferences of real lepton-quark(\& gluon) scattering amplitudes describing the radiation of a photon emitted by either the lepton or the quark. Such real contributions typically contain phase space integrations. Interferences from virtual two-photon-exchange one-loop diagrams and single photon exchange diagrams may also contribute. The various hard cross sections can be combined by application of QCD-EOM inspired relations between effects of transverse partonic motion and multipartonic correlations, and eventually the soft divergences indicated by poles in $1/\varepsilon$ cancel~\cite{SameQuark}.\\
The hard cross sections can be computed along the same lines for scenario (2). To leading order only tree-level diagrams interfere without phase space integrations~\cite{DiffQuark}. Hence, no soft divergences appear in intermediate steps of the calculation.\\
Adding the results of Refs.~\cite{SameQuark,Afanasev,DiffQuark} leads to the following parton picture formula for the single transverse spin dependent DIS cross section at $\mathcal{O}(\alpha_{\mathrm{em}}^3)$,
\begin{eqnarray}
E^\prime \frac{\d\sigma_{UT}}{\d^3l^\prime} & = & -|S_T|\sin\phi_s \frac{4\alpha_{\mathrm{em}}^3}{yQ^4} \frac{M}{Q}\frac{x_B y}{\sqrt{1-y}}\times\label{eq:result}\\
& &\sum_q \Bigg[ e_q^3\int_0^1dx\Big(\hat{C}_+(x,x_B,y)\ G_F^q(x_B,x)+\hat{C}_-(x,x_B,y)\ \tilde{G}_F^q(x_B,x)\Big)\nn
& & +e_q^3(1-y)\frac{m_q}{M}h_1^q(x_B)+\frac{2-y}{2y}e_q^2(1-x_B\frac{\d}{\d x_B}) G_F^{\gamma,q}(x_B,x_B)\Bigg].\nnn
\end{eqnarray}
The perturbative coefficient functions $\hat{C}_\pm$ in Eq.~(\ref{eq:result}) are integrable distributions and their functional form is given in Ref.~\cite{SameQuark}. Assuming that the non-perturbative quark-gluon correlation functions $G_F^q(x,x^\prime)$ and $\tilde{G}_F^q(x,x^\prime)$\footnote{Definitions in terms of hadronic matrix elements for both functions can be found in Ref.~\cite{ABMS}.} are analytic the $x$-integral in (\ref{eq:result}) is well-defined. In addition the finite quark mass term of Ref.~\cite{Afanasev,DiffQuark} has been added to Eq.~(\ref{eq:result}) as well as the contribution of Ref.~\cite{DiffQuark} describing scenario (ii) where the two photons couple to different quarks. The latter term involves a quark-photon correlation function $G_F^\gamma$\footnote{Notice a slight redefinition $G_F^\gamma(x,x)\equiv\frac{1}{2e^2}F_{FT}(x,x)$ of the object $F_{FT}$ introduced in \cite{DiffQuark}.}.\\
The SSA $A_{UT}$ can be computed from (\ref{eq:result}) in the following way ($\d\sigma=E^\prime \d\sigma/\d^3k^\prime$),
\begin{equation}
 A_{UT}=\frac{\d\sigma_{UT}(\phi_s)-d\sigma_{UT}(\phi_s-\pi))}{2\d\sigma_{UU}}\ ,\label{AUT}
\end{equation}
with the well-known parton model result for the unpolarized cross section~\cite{Baccetal},
\begin{equation}
\d\sigma_{UU}  = \frac{4\alpha_{\mathrm{em}}^2}{Q^4y}f(y)\sum_qe_q^2x_B f_1^q(x_B)\ ,\label{UU}
\end{equation}
with $f_1$ the unpolarized collinear parton distribution.
\section{Model for the quark-gluon correlations from light cone wave functions}
In order to utilize Eq.~(\ref{eq:result}) to estimate the sign and size of the transverse target spin asymmetry $A_{UT}$ on a proton and neutron one needs information on the full support of the non-perturbative quark-gluon correlation functions $G_F^q(x,x^\prime)$, $\tilde{G}_F^q(x,x^\prime)$\footnote{Note that $G_F^q(x,x^\prime)=G_F^q(x^\prime,x)$ and $\tilde{G}_F^q(x,x^\prime)=-\tilde{G}_F^q(x^\prime,x)$. Hence, $\tilde{G}_F^q(x,x)=0$.}, as well as the quark-photon correlation function $G_F^{\gamma,q}(x,x)$ in the soft photon limit $x^\prime = x$ and the transversity distribution $h_1^q(x)$. Currently, only extractions from data exist for the so-called "Soft Gluon Pole matrix element" $G_F^q(x,x)$~\cite{Param} and the transversity distribution. However, a recent model calculation gives predictions for $G_F^q$ and $\tilde{G}_F^q$ on the full support $x\neq x^\prime$~\cite{Braun}. In this work the twist-3 quark-gluon correlation functions are expressed in terms on non-valence-like light cone wave functions, and analytical results at a scale $\mu_0 = 1 \ \mathrm{GeV}$ were obtained. One specific feature of this model is that $G_{F,\mathrm{Model}}^q(x,x,\mu_0)=0$ due to the absence of final state interactions. This is in obvious contradiction to parametrizations from data for $G_F^q(x,x)$~\cite{Param}, and the model does not properly describe the physics of $G_F^q(x,x^\prime)$ in a small interval around $x^\prime \sim x$. Nevertheless it may realistically probe the physics outside of this interval, i.e., where $x^\prime$ is further away from $x$. Under the approximation that the quark-photon matrix element $G_F^{\gamma,q}$ is proportional to the quark-gluon matrix element $G_F^q$~\cite{DiffQuark} one also has $G_{F,\mathrm{Model}}^{\gamma,q}(x,x,\mu_0)=0$.
\begin{figure}
\centering
\includegraphics[width=7cm]{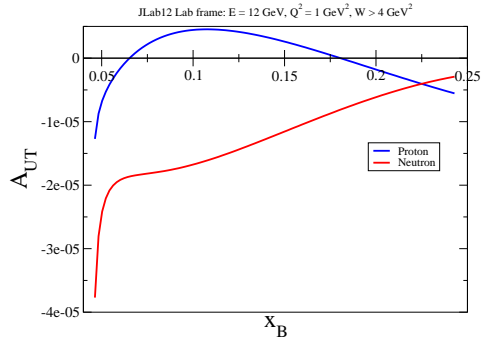} 
\caption{Prediction for the asymmetry $A_{UT}$ at JLab12 obtained from a model~\cite{Braun}. \label{plot}}
\end{figure}
Hence, for massless quarks the asymmetry $A_{UT}$ in (\ref{AUT}) is completely determined by $G_{F,\mathrm{Model}}^{u,d}(x,x^\prime,\mu_0)$ and $\tilde{G}_{F,\mathrm{Model}}^{u,d}(x,x^\prime,\mu_0)$ at a fixed scale $Q=1\ \mathrm{GeV}$. The model prediction for $A_{UT}$ is shown in Fig.~\ref{plot}. In this plot fixed target kinematics have been used for an electron beam energy $E=12\ \mathrm{GeV}$ (JLab12 kinematics). A missing mass $W=(P+q)^2>4\ \mathrm{GeV^2}=W_{\mathrm{min}}$ was assumed to ensure that the asymmetry is probed in the DIS region. This defines a maximal Bjorken-$x$ $x_{B,\mathrm{max}}=Q^2/(Q^2+W_{\mathrm{min}}-M^2)\sim 0.25$ for $Q=1\ \mathrm{GeV}$. For a fixed scale the energy transfer from the electron to the nucleon $y$ varies with $x_B$, that is, $y=Q^2/(2MEx_B)$. Typical experimental values $y\sim 0.4-0.6$ are probed at $xB\sim 0.1$ at $Q=1\ \mathrm{GeV}$.

\section{Conclusions}
The plot in Fig.~\ref{plot} shows that one can expect rather small asymmetries of about $10^{-5}$ from the model of Ref.~\cite{Braun}. Although the JLab data \cite{JLab} for the SSA $A_{UT}$ on a neutron is still preliminary it gives hints that the asymmetry is much larger in reality for a neutron. This discrepancy may point to missing physics in the integration region $x\sim x^\prime$ in Eq.~(\ref{eq:result}) which is left out in the model of Ref.~\cite{Braun}. One may consider larger values of $Q>1\ \mathrm{GeV}$. At larger scales a non-zero "Soft Gluon Pole" $G_F^q(x,x,\mu>1\ \mathrm{GeV})\neq 0$ can be obtained from evolution of the model results of Ref.~\cite{Braun}. However, one would not expect the asymmetry to be dramatically larger at higher scales due to the factor $M/Q$ in Eq.~(\ref{eq:result}).


\end{document}